# Traditional aging theories: which ones are useful?

Jicun Wang-Michelitsch[1]*, Thomas M Michelitsch[2]

[1]Department of Medicine, Addenbrooke's Hospital, University of Cambridge, UK (Work address until 2007)

[2] Institut Jean le Rond d'Alembert (Paris 6), CNRS UMR 7190 Paris, France

**Abstract**

Many theories have been proposed for answering two questions on aging: "Why do we age?" and "How do we age?" But which ones of them are finally tenable? We have made an analysis of these theories to select out the useful ones. Evolutionary theories are made for interpreting the evolutionary advantage of aging, and "saving resources for group benefit" is thought to be the purpose of aging. However, for saving resources, a more economic strategy should be to make the individuals over reproduction age undergo a rapid death rather than undergo a slow aging. Biological theories are made for identifying the causes and the biological processes of aging. However, some of them including cell senescence/telomere theory, gene-controlling theory, and developmental theory, have unfortunately ignored the influence of damage on aging. Free-radical theory suggests that free radicals by causing intrinsic damage are the main cause of aging. However, even if intracellular free radicals can cause injuries, they may be only associated with some, but not all, of the aging changes. Damage (fault)-accumulation theory predicts that the intrinsic faults as damage can accumulate and lead to aging.  But the reality is that, an unrepaired fault cannot possibly remain in a living organism, since it can destroy the structural integrity of a tissue and cause a rapid failure of the organism. Therefore, these theories are untenable on interpreting aging. Nevertheless, among them, the developmental theory and the damage (fault)-accumulation theory are more useful than others, because they emphasize the importance of damage and body-development in aging. Some physical theories are also useful by pointing out the common characteristics of aging process, such as loss of complexity, increase of entropy, and failure of information-transmission. An advanced aging theory should be a theory that can include all of the useful ideas in these traditional theories.

"Why do we age?" and "How do we age?" are two different questions. For the first, we need to explain why aging is unavoidable; and for the second, we need to explain the underlying biological process of aging. Many theories have been created for answering these two questions. Evolutionary theories, including evolvability theory (Weismann, 1889), mutation-accumulation theory (Medawar, 1952), antagonistic pleiotropy theory (Williams, 1957), and disposable soma theory (Kirkwood, 1977), are proposed to interpret the evolutionary advantage of aging. Biological theories are made to identify the causes and the biological processes of aging. For example, free-radical theory emphasizes the central role of free radicals in aging (Harman, 1956). Cell senescence/telomere theory suggests that the telomere-controlled senescence of cells is the main cause of aging (Blackburn, 2000). Developmental theory supposes that aging is a result of development (Medvedev, 1990). Damage (fault)-accumulation theory and gene-controlling theory are actually the leading theories nowadays. Damage (fault)-accumulation theory predicts that intrinsic faults, which are left unrepaired due to the limitation of repair/maintenance, are the sources of aging (Kirkwood, 2005 and 2006). Gene-controlling theory demonstrates that it is certain genes that control the process of aging completely and independently. Physical theories have summarized the physical properties of aging, such as loss of complexity (Lipsitz, 1992), consequence of increase of entropy (Bortz, 1986), and failure of information-transmission (Mulá, 2004). Now the question is, among all of these theories, which ones are finally tenable in explaining aging? In the present paper, we will analyze each of them to answer this question. Our discussion is devoted to the following subjects:

I. A method for disproving a theory

II. Limitations of traditional aging theories

    2.1 Evolutionary theories
    2.2 Free-radical theory
    2.3 Cell senescence/telomere theory
    2.4 Gene-controlling theory
    2.5 Developmental theory
    2.6 Damage (fault)-accumulation theory
    2.7 Physical theories

III. Useful theories and unanswered questions

I. **A method for disproving a theory**

To accept or abandon a theory, it is important to find out negative evidence rather than to collect positive evidence. A theory can be falsified by even "one" negative fact (Popper, 1994). When we search for a real criminal from a number of suspects, a feasible and simple way is to exclude at first the impossible suspects. On judging a theory, this strategy is also useful. We can analyze the impossibility of a theory by collecting negative evidence. On this basis, we propose here a method for disproving a theory, which we name as "deductive analysis of impossibility". Taken the gene-controlling theory as an example, the procedure of analysis is as follows: **Premise 1**, if the prediction that "aging is controlled by a gene" is correct, the distribution of aging changes should be independent of the locations of damage (Figure 1A); **Premise 2**, in fact, the distribution of wrinkles on the face is closely related to the locations of damage (Figure 1A and Figure 1B); and **Conclusion**, therefore the prediction of "aging is controlled by a gene" is impossible, and gene-controlling theory is untenable (Figure 1A). When a negative fact exists, a theory is not any more acceptable; despite it has numbers of positive evidence. Unfortunately, some biologists have forgotten this principle in judging their works. They make all the efforts to collect positive data, but their conclusion can be disproved by a single negative case. In the following discussion, we will analyze these aging theories by this method.

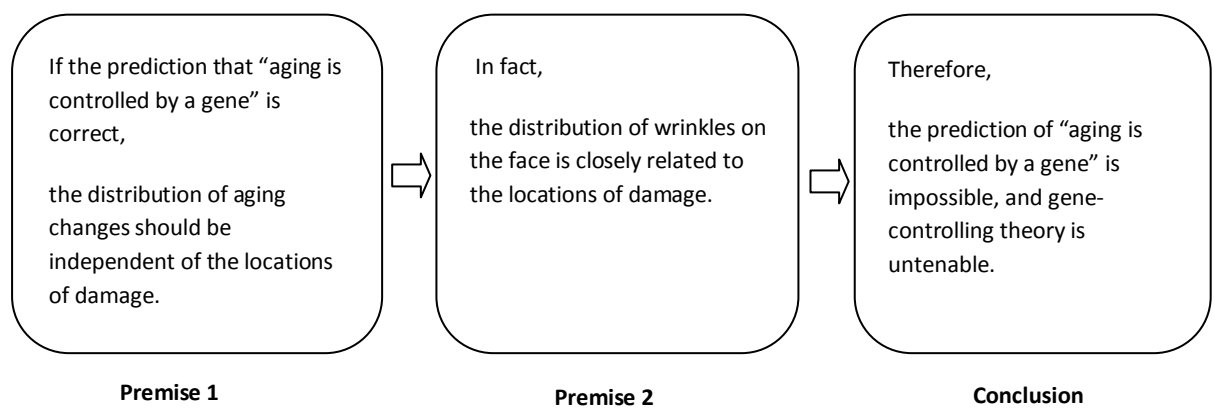

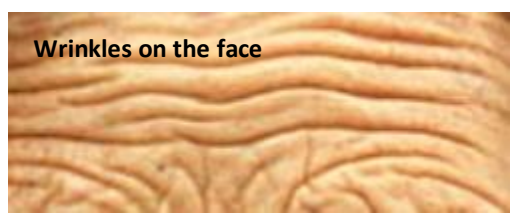

**Figure 1. Deductive analysis of impossibility on an aging theory**

"Deductive analysis of impossibility" is a useful method to disprove an aging theory. We can analyze the impossibility of a theory by collecting negative evidence. Taken the gene-controlling theory as an example, the process of deductive analysis is as follows: **Premise 1**, if the prediction that "aging is controlled by a gene" is correct, the distribution of aging changes should be independent of the locations of damage (**A**); **Premise 2**, in fact, the distribution of wrinkles on the face is closely related to the locations of damage (**B**); and **Conclusion**, therefore the prediction of "aging is controlled by a gene" is impossible, and the gene-controlling theory is untenable (**A**).

## II. Limitations of traditional aging theories

Traditional aging theories interpret aging from different aspects; however, so far, none of them can give a full story of aging mechanism. In this part, we will look into each of these theories to tell why they are not complete and why some of them are not at all acceptable. An untenable theory should be given up. However, the useful ideas in it should be inherited, because they are important for further studying of aging.

## 2.1 Evolutionary theories

One of the earliest theories on aging is the evolvability theory, which was firstly proposed by Weismann but recently modified by Mitteldorf (Weismann, 1889; Mitteldorf, 2006). This theory suggests that the purpose of limited lifespan is to save living resources for younger generations for "group benefit". In this direction, some theories provided some possible mechanisms on how an organism "sacrifices" itself for "group benefit". These theories include mutation-accumulation theory, antagonistic pleiotropy theory, and disposable soma theory. It is true that group benefit has an evolutionary advantage. However, for saving resources for reproduction, a more effective strategy should be to remove the old individuals through "a planned rapid death" rather than through a long process of aging. In reality, the maximum lifespans of individuals of most species' of animals are much longer than the time for development. For example, a woman can live more than 30 years (age 70-90) after reproduction age (age 15-45).

**Mutation-accumulation theory:** The main point of mutation-accumulation theory is that weak fitness-selection on an old organism allows a wide range of gene mutations in somatic cells, which have deleterious effects on the organism (Medawar, 1952; Martin, 1996). Those DNA mutations that take place after reproduction age cannot affect next generations. Therefore, such mutations cannot be wiped out by evolution, and they can accumulate in a somatic body. However, in reality, the same type of aging changes such as age spots often develop parallelly and independently in different parts of an organ. The aging changes that develop in this way cannot be possibly a result of DNA mutations. How is it impossible that the same form of DNA change takes place parallelly and independently in different somatic cells in different parts of an organ? Another question over this theory is: why are the cells with altered phenotypes by DNA changes not removed by the immune system if they do not develop into tumors?

**Antagonistic pleiotropy theory:** In 1957, Williams gave another supposition on aging symptoms in his antagonistic pleiotropy theory (Williams, 1957). He supposed that some genes with antagonistic pleiotropic effects might exist, which have benefiting effects in early life but detrimental effects in later age. Williams also pointed out that ageing is a side effect of necessary functions and any alteration of ageing process is "impossible." This theory sounds reasonable by presuming a kind of pleiotropic genes. However, such genes cannot possibly exist, because their detrimental effects in all somatic cells should result in a quick death rather than a gradual aging of organs. In reality, our organs can maintain its sufficient functionality for more than 60 years even if they are aging!

**Disposable soma theory:** Disposable soma theory presumes that an organism must budget the limited energy available to it and the energy has to be well distributed for metabolism, reproduction, and repair/maintenance. An organism prefers to save energy for reproduction rather than maintain the mature body. Insufficient repair is therefore the cause for deleterious changes of body with age (Kirkwood, 1977). However, insufficiency of repair can be fatal for an organism. If a wound on skin is left unrepaired due to insufficiency of repair, the consequence would be death of the body from bleeding or infection rather than aging of the skin.

## 2.2 Free-radical theory

This theory points out that free radicals are the main origin of aging by causing cellular injuries (Harman, 1956). Molecules of free radicals in biological organisms include superoxide and the reactive oxygen species. These small molecules are produced in cells as substrates or by-products in some biochemical reactions. Although some studies have shown the effects of anti-free-radical on extending lifespans of animals such as Drosophila and roundworms (Van Raamsdonk, 2009), the role of free radicals in aging is controversial (Ahluwalia, 2004; Juránek, 2005). Oxidative free radicals can cause potentially cell injuries if they are in a free state. However, as part of cell metabolism, these molecules should be in an accurate control on distributions and on activities like other molecules in cells. Normally free radicals will be immediately removed or deactivated after functioning by intracellular antioxidants for maintaining intracellular stability. Even if they cause sometimes injuries, the injured cells or molecules can be fully repaired or removed, not necessarily affecting tissue structure and tissue function. Even if some injuries made by free radicals cannot be fully repaired, these injuries are associated only with some, **but not all,** of the aging changes.

## 2.3 Cell senescence/telomere theory

Reduction of cell number is always thought to be a cause of aging. This idea is enhanced by the discovery of telomeres. Telomeres are the ending parts of chromosomes, and they are important in protecting chromosomes and in regulating the activities of chromosomes. During DNA duplication, the 3' ending part of a template DNA in telomere is used as the priming sequence for synthesizing new DNA. Therefore, the new DNA is shorter than template DNA, and the telomere in new DNA is shorter than that in template DNA. The length of a telomere is therefore shortened after each time of DNA duplication and cell division. When a telomere is too short, DNA synthesis and cell division cannot anymore proceed. On this basis, a cell senescence/telomere theory was proposed. This theory suggests that the telomere-determined cell senescence is the origin of aging (Hayflick, 1965; Blackburn, 2000). The discovery of telomerase, an enzyme that can prolong telomeres, was a powerful support to this theory, because telomerase was found to be produced mainly in germ cells and in tumor cells (Kim, 1994). However, more and more studies have shown that there is no consistent relationship between telomere lengths and animal lifespans (Kipling, 1990; Hemann, 2000). In addition, this theory is incompatible with some known facts: **A.** the potential of cell division is also controlled by other factors, and a stem cell does not essentially have longer telomeres than its neighbor non-stem cells; **B.** reduction of cell number is not the unique change in an aged tissue, and in contrast some aging changes including hyperplasia appear as increase of number of cells; and **C.** most of our organs can still reproduce new cells for repair when we are 70 years old.

## 2.4 Gene-controlling theory

Life science is nowadays in an era of "gene" and "genome". Aging is thought to be solvable by discovering the controlling genes. We summarize this belief as a gene-controlling theory. Searching for "aging-related genes" and "lifespan-related genes" becomes therefore the main work for many scientists (de Magalhaes, 2009; Fabrizio, 2010; Bell, 2012; McCormick, 2012). The belief of gene-controlling on aging has the same fundamental as that of gene-programming theory, in which aging process is thought to be pre-programmed genetically as that in development (Longo, 2005). The earliest idea of programmed aging might come from the phenomenon of planned death (biological suicide) in some species' of animals. Individuals of semelparous salmon and female octopus go to die quickly following reproduction. Hormone signaling has been identified to be the mechanism for the programmed death of octopus (Wodinsky, 1977). However, this programmed death is sudden and short, quiet different from that of aging process. In reality, the individuals of the same species but living in different environments, protected or not protected, may have quite different lifespans. Even if in the same environment, the individuals die at different ages. The idea of programmed aging has been given up by most scientists (Austad, 2004).

Many aging-related genes and lifespan-related genes have been identified (McCormick, 2012; University of Washington, 2012); however their exact roles in aging are not known. In our view, the belief in aging genes is misleading, because this idea has severe defects. **A**. Even if some genes are associated with lifespans, they are not essentially aging-related. **B**. Even if some genes are aging-related, they are not essentially the elements that control aging process completely and independently. **C**. If aging is controlled by certain genes, the distribution of aging changes should be independent of the locations of damage. However, the fact is that the distribution of age spots on the skin is closely related to the locations of damage. **D**. If aging is controlled completely by a gene, when this gene fails in an individual by DNA mutations, this individual could be immortal. However, no immortal case has been observed so far in the world. Therefore, such a gene and such a programming mechanism for aging may not possibly exist!

## 2.5 Developmental theory

Alternation of development pathway can alter significantly the lifespan of an organism (Klass, 1976). On this basis, Johnson assumed that aging and development are coupled (Johnson et al, 1984), and Medvedev proposed his developmental theory. This theory suggests that aging is a result of development, and aging process is regulated by the same mechanism as that in development (Medvedev, 1990; Zwaan, 2003). This idea was supported by the discovery of correlation between the potential of longevity and the mature time of an animal (Charnov, 1993). *Dauer* pathway is an altered pathway of development of *C. Elegans,* and *Dauer* individuals have prolonged lifespans. *Dauer* pathway was found to be induced by an environment change and be mediated by insulin-like signaling pathway (Wolkow, 2000).

Hormones and hormone-related genes were then thought to be important in regulating aging process apart from regulating development (de Magalhaes, 2005). The idea of coupling the process of aging with the process of development is revealing; however this theory has three defects: **A**. it has ignored the influence of damage on aging; **B**. it does not clarify the difference between development process and aging process; and **C**. the effect of alteration of development on lifespan is not essentially via retarding aging.

### 2.6 Damage (fault)-accumulation theory

Somatic injuries are un-negligible in development of aging changes. In this aspect, Kirkwood predicted that: *"aging is a result of accumulation of 'faults' at cellular and molecular level because of the limitation of maintenance and repair; the underlying driving force is damage. The genetic control of longevity comes through the regulation of the essential maintenance and repair processes that slow the build–up of faults"* (Kirkwood, 2005). We summarize this prediction as "damage (fault)-accumulation theory". This theory emphasizes the importance of damage and repair/maintenance system in aging. However, since some terms are not specified, the conclusion of this theory is confusing. The term of "fault" in this prediction can be understood as a kind of "intrinsic damage". However, it is not clear whether this "intrinsic damage" is referred to a change of a molecule/cell before repair or a change after repair. For example, in scar formation of burnt skin, the wound of skin by burning is the primary damage before repair, and the scar is the change of skin after repair. These two changes of skin before repair and after repair are obviously different; thus they should be distinguished. The term of "faults" in this prediction is more possibly referred to primary damage before repair, since "they accumulate because of the limitation of repair/maintenance". A critical question over this theory is: which kind of "faults" can be left unrepaired. In our view, if the faults are a kind of primary damage, they cannot remain unrepaired in a living organism. Primary damage is in fact a defect of a living structure such as molecule, cell, or tissue. If the defect is not closed in time, the structure will lose its structural integrity and functionality, and the whole organism will fail rapidly. In another word, unrepaired faults will lead to a rapid death rather than aging of an organism. Hence, the concept of accumulation of "faults" is misleading, and the main idea of this theory is untenable.

### 2.7 Physical theories

Aging does not only take place in living organisms, but also in some non-living structures. Searching for the common as well as the different characteristics of aging on these two different systems is important for uncovering the secret of aging. Some physical theories give objective descriptions of aging in living- and non-living systems, such as loss of complexity, increase of entropy, and failure of information-transmission. These physical theories are therefore useful and revealing. But a physical theory cannot give a full story of aging mechanism, because it cannot describe the biological process of aging.

**Loss of complexity:** A living system is a kind of complex adaptive system that is composed of numbers of interacting sub-systems. The term of complexity is an approach in physics to describe the interactions between a whole system and its sub-systems. Complexity is in ratio to the degree of difficulty in predicting the properties of the whole system given the properties of its sub-systems (Weaver, 1948). The complexity of a system exhibits on several aspects, for example, emergence, feedback effect (interdependence), self-organization, and self-adaptation (Morowitz, 1995). Loss of complexity appears as failure of communications between sub-systems. In a living system, loss of complexity appears as degeneration of cells/tissues and chaos of communications between cells/tissues (Lipsitz, 1992). In the loss-of-complexity (chaos) theory, Lipsitz pointed that it is the loss of complexity with loss of functionality that occurs as aging. Aging is thus a price of cooperation and complexity (Kiss, 2009).

**Increase of entropy:** Entropy is a useful measure for marking the tendency of natural processes, which are always in a direction of increase of entropy in a closed system. The term of entropy is also understood as "uncertainty" or "lack of information" in some fields (MacKay, 2003). The statistic interpretation of entropy was firstly introduced by Boltzmann as a phenomenological measure of "disorder" (Landsberg, 1988). Increase of disorder is corresponding to the decrease of complexity and functionality of a system. Living systems need to release entropy continuously to protect them from running into thermodynamic equilibrium and death (Niedermueller, 1990; Haken, 1990; Wunderlin, 1992). Respiration, perspiration, and excretion are the measures to release entropy. However, a living being cannot get rid of all of the entropy, and part of it will remain and increase with age. Aging is therefore a consequence of increase of entropy. Entropy increases when the energies for ordering diminish (Bortz, 1986).

**Failure of information transmission:** In the theory of Multi-cellular Being Chaos, it is emphasized that the multi-cellular structure of an organism based on cell differentiation is a precondition for aging (Mulá, 2004). The information-transmission pathways between different cells are the surviving lines for cells and for the organism. With time, signal transmissions become chaos gradually; and that is the nature of aging. A colony of bacterial has no aging, because in a colony there is neither cell organization nor information transmission between cells. Failure of information transmission is corresponding to the loss of complexity of an organization. This idea is outstanding on recognizing the importance of cell differentiation and cell organization in aging.

### III. Useful theories and unanswered questions

Our analysis shows that these theories are incomplete and an advanced theory is needed. A better theory should be consistent with the facts that old theories have uncovered. These facts include: **A**. Damage is the driving force for aging (in the damage (fault)-accumulation theory); **B**. Modification of development can alter lifespan (in the developmental theory); and **C.** Aging is a process of loss of complexity and failure of information transmission (in the physical theories). Therefore, among all these theories, damage (fault)-accumulation theory, developmental theory, and physical theories are more useful than others. So far, some basic questions over aging are not yet satisfactorily answered, and they are: **A.** why aging is unavoidable; **B.** how we age; and **C.** why we have limited longevity. Therefore, a better theory should also be able to answer these critical questions. To this end, we have proposed a novel theory on aging, the Misrepair-accumulation theory (Wang, 2009).

**References**

1. Ahluwalia J, Tinker A, Clapp LH, Duchen MR, Abramov AY, Pope S, Nobles M, and Segal AW. (2004). The large-conductance Ca2+-activated K+ channel is essential for innate immunity. Nature. 427: 853–858 (2004). doi:10.1038/nature 02356
2. Amrit FRG, Boehnisch CML, May RC. (2010) Phenotypic Covariance of Longevity, Immunity and Stress Resistance in the Caenorhabditis Nematodes. PLoS ONE. 5 (4): e9978
3. Anderson, RM, Shanmuganayagam D, Weindruch R. (2009). Caloric Restriction and Aging: Studies in Mice and Monkeys. Toxicologic Pathology. 37 (1): 47–51
4. Austad, S. N. (2004). Is aging programed? Aging Cell 3(5): 249-251
5. Bak P. (1996). How Nature Works: The Science of Self-Organized Criticality, Copernicus press, New York, U.S. Cloth: ISBN 0-387-94791-4
6. Bansal D, Miyake K, Vogel SS, Groh S, Chen CC, Williamson R, McNeil PL, and Campbell KP. (2003). Defective membrane repair in dysferlin-deficient muscular dystrophy. Nature. 423: 168
7. Bell JT, Tsai P-C, Yang T-P, Pidsley R, Nisbet J, et al. (2012). Epigenome-Wide Scans Identify Differentially Methylated Regions for Age and Age-Related Phenotypes in a Healthy Ageing Population. PLoS Genet. 8(4): e1002629. doi:10.1371/journal.pgen.1002629
8. Bishay K, Ory K, Olivier MF, Lebeau J, Levalois C, Chevillard S. (2001). DNA damage-related RNA expression to assess individual sensitivity to ionizing radiation. Carcinogenesis. 22(8): 1179
9. Blackburn EH. (2000). Telomere states and cell fates. Nature. 408: 53-56
10. Bodnar AG, Ouellette M, Frolkis M, Holt SE, Chiu CP, Morin GB, Harley CB, Shay JW, Lichtsteiner S, and Wright WE. (1998). Extension of life-span by introduction of telomerase into normal human cells. Science. 279: 349-352
11. Bortz WM 2nd. (1986). Aging as entropy. Exp Gerontol. 21(4-5): 321-8
12. Bulckaen H, Prévost G, Boulanger E, et al. (2008) Low-dose aspirin prevents age-related endothelial dysfunction in a mouse model of physiological aging. Am J Physiol Heart Circ Physiol. 294: 1562
13. Charnov EL. (1993). Life History Invariants: Some Explorations of Symmetry in Evolutionary Ecology. Oxford University Press, Oxford
14. Collins FS, et al. (2004). International Human Genome Sequencing Consortium. Finishing the euchromatic sequence of the human genome. Nature. 431: 931-945 | doi:10.1038/nature03001
15. Cook H, Stephens P, K Davies J, Harding KG and Thomas DW. (2000). Defective Extracellular Matrix Reorganization by Chronic Wound Fibroblasts is Associated with Alterations in TIMP-1, TIMP-2, and MMP-2 Activity. Journal of Investigative Dermatology. 115: 225
16. de Magalhães JP and Church GM. (2005). Genomes optimize reproduction: aging as a consequence of the developmental program. Physiology. 20:252-259
17. de Magalhaes JP, Curado J, and Church GM. (2009). Meta-analysis of age-related gene expression profiles identifies common signatures of aging. Bioinformatics. 25: 875-881
18. Fabrizio P, Hoon S, Shamalnasab M, Galbani A, Wei M, Giaever G, Nislow C, Longo VD. (2010). Genome-wide screen in Saccharomyces cerevisiae identifies vacuolar protein sorting, autophagy, biosynthetic, and tRNA methylation genes involved in life span regulation. PLoS genetics. 6(7):e1001024.
19. Haken H and Wunderlin A. (1990). Synergetik: Eine Einführung Nichtgleichgewichts-... und Selbstorganisation in Physik, Chemie und Biologie. In: Kratky KW. und Wallner F. Grundprinzipien der Selbstorganisation / Hrsg. - Darmstadt : WBG, (1990). - ISBN 3-534-10971-6.
20. Harman D. (1956). Aging: a theory based on free radical and radiation chemistry. Journal of Gerontology. 11 (3): 298–300
21. Hemann MT; Greider, CW. (2000). Wild-derived inbred mouse strains have short telomeres. Nucleic Acids Research. 28 (22): 4474–4478

22. Herndon LA, Schmeissner PJ, Dudaronek JM, Brown PA, Listner KM, Sakano Y, Paupard MC, Hall DH, Driscoll M. (2002). Stochastic and genetic factors influence tissue-specific decline in ageing C. elegans. Nature. 419: 808–814
23. Hayflick L. (1965). The limited in vitro lifetime of human diploid cell strains. Exp Cell Res. 37: 614-636
24. Johnson TE, Mitchell DH, Kline S, Kemal,R and Foy J. (1984). Arresting development arrests aging in the nematode Caenorhabditis elegans. Mech Ageing Dev. 28 (1):23-40
25. Johnson S. (2001). Emergence: the connected lives of ants, brains, cities, and software. New York: Scribner. ISBN 0-684-86875-X.
26. Juránek I. and Bezek Š. (2005). Controversy of Free Radical Hypothesis: Reactive Oxygen Species – Cause or Consequence of Tissue Injury? Gen. Physiol. Biophys. 24: 263—278
27. Kim NW, Piatyszek MA, Prowse KR, Harley CB, West MD, Ho PL, Coviello GM, Wright WE, Weinrich SL, and Shay JW. (1994). Specific association of human telomerase activity with immortal cells and cancer. Science. 266: 2011-2015
28. Kipling D and Cooke HJ. (1990). Hypervariable ultra-long telomeres in mice. Nature. 347 (6291): 400
29. Kirkwood TB. (1977). Evolution of ageing. Nature. 270 (5635):301–304
30. Kirkwood TB and Austad SN. (2000). Why do we age? Nature Genet. 408 (6809): 233-8
31. Kirkwood TB. (2006). Ageing: too fast by mistake. Nature. 444 (7122): 1015-7
32. Kirkwood TB. (2005). Understanding the odd science of aging. Cell. 120(4): 437-47
33. Kiss HJM, Mihalik A, Nanasi T, Ory B, Spiro Z, Soti C, and Csermely P. (2009). Ageing as a price of cooperation and complexity: Self-organization of complex systems causes the ageing of constituent networks. BioEssays. 31(6): 651-64. arXiv:0812.0325v2 [q-bio.MN] 01/2009
34. Klass M and Hirsh D. (1976). Non-ageing developmental variant of Caenorhabditis elegans. Nature. 260(5551):523-5
35. Landsberg PT. (1984). Can Entropy and "Order" Increase Together? Physics Letters. 102A: 171–173
36. Lipsitz LA, Goldberger AL. (1992). Loss of 'complexity' and aging. Potential applications of fractals and chaos theory to senescence. JAMA. 267(13): 1806-9
37. Lipsitz LA. (2006). Aging as a Process of Complexity Loss, Complex Systems Science in Biomedicine, Topics in Biomedical Engineering. International Book Series, Part III, Section 7, 641-654, DOI: 10.1007/978-0-387-33532-2_28
38. Li, TY and Yorke JA. (1975). Period Three Implies Chaos. American Mathematical Monthly 82 (10): 985
39. Longo VD, Mitteldorf J, and Skulachev VP. (2005). Programmed and altruistic ageing. Nat Rev Genet. 6 (11):866-872
40. MacKay, DJC. (2003) Information Theory, Inference, and Learning Algorithms Cambridge: Cambridge University Press. 2003. ISBN 0-521-64298-1
41. Martin GM, Austad SN and Johnson TE. (1996). Genetic analysis of aging: role of oxidative damage and environmental stresses. Nature Genet. 13 (1):25-34
42. May RM. (1976). Simple mathematical models with very complicated dynamics. Nature. 261 (5560): 459–467.doi:10.1038/261459a0. PMID 934280.
43. McCormick M, Chen K, Ramaswamy P, and Kenyon C. (2012). New genes that extend Caenorhabditis elegans' lifespan in response to reproductive signals. Aging Cell. 11(2):192-202
44. Medawar PB. (1952). An Unsolved Problem of Biology. Lewis, London
45. Medvedev ZA. (1990). An attempt at a rational classification of theories of ageing. Biol Rev Camb Philos Soc. 65 (3): 375-398
46. Mitteldorf J. (2006). Chaotic population dynamics and the evolution of ageing: proposing a demographic theory of senescence. Evol Ecol Res. 8: 561–74
47. Morowitz HJ and Singer JL. (eds.1995) The Mind, the Brain, and Complex Adaptive Systems. Addison-Wesley ISBN. 0201409860
48. Mulá MAV. (2004). Why do we age? http://www.meucat.com/vi.html.
49. Niedermueller H and Hofecker G. (1990). Anwendungsmöglichkeiten des Prinzips der Autopoiese auf physiologische und gerontologische Fragestellungen. In: Kratky KW. und Wallner F. Grundprinzipien der Selbstorganisation / Hrsg. - Darmstadt : WBG, (1990). - ISBN 3-534-10971-6.

50. Popper KR. (1994). Zwei Bedeutungen von Falsifizierbarkeit [Two meanings of falsifiability]. In Seiffert H and Radnitzky G. Handlexikon der Wissenschaftstheorie. München: Deutscher Taschenbuch Verlag. pp. 82–85. ISBN 3-423-04586-8.
51. University of Washington Aging Genes and Interventions Database (2012): http://www.primateportal.org/link/university-washington-aging-genes-and-interventions-database
52. Van Raamsdonk JM, Hekimi S. (2009). Deletion of the Mitochondrial Superoxide Dismutase sod-2 Extends Lifespan in Caenorhabditis elegans. PLoS Genetics. 5 (2): e1000361.
53. Wang J, Michelitsch T, Wunderlin A, Mahadeva R. (2009) "Aging as a Consequence of Misrepair – a Novel Theory of Aging". ArXiv: 0904.0575. arxiv.org
54. Weaver W. (1948). Science and Complexity. " American Scientist. 36 (4): 536–44
55. Weismann A. (1889). Essays upon heredity and kindred biological problems. Clarendon Press, Oxford
56. Williams GC. (1957). Pleiotropy, natural selection, and the evolution of senescence. Evolution. 11:398–411
57. Wodinsky J. (1977). Hormonal Inhibition of Feeding and Death in Octopus: Control by Optic Gland Secretion. Science. 148 (4320): 948–51
58. Wolkow CA, Kimura, KD, Lee MS, and Ruvkun G. (2000). Regulation of C. elegans life-span by insulinlike signaling in the nervous system. Science. 290: 147–150
59. Wunderlin A and Friedrich R (1992). Evolution of Dynamical Structures in Complex Systems. Springer-Verlag, Berlin. ISBN 3642847838
60. Zwaan, B. (2003). Linking development and aging. Sci Aging Knowledge Environ. 47: 32